\documentclass[aps,prb,12pt]{revtex4-1}

\usepackage{amsmath}
\usepackage{amsfonts}
\usepackage{amssymb}
\usepackage{graphicx}
\usepackage{color}
\usepackage[colorlinks=true,citecolor=red,linkcolor=blue]{hyperref}

\def\mathbi#1{\textbf{\em #1}}

\begin{document}

\title{Molecular plasmonics: the role of ro-vibrational molecular states in exciton-plasmon materials under strong coupling conditions}

\author{Maxim Sukharev}
\affiliation{Arizona State University, Mesa, Arizona 85212, USA}
\email{maxim.sukharev@asu.edu}

\author{Eric Charron}
\affiliation{Institut des Sciences Mol\'eculaires d'Orsay (ISMO), Univ. Paris-Sud, CNRS, Universit\'e Paris-Saclay, 91405 Orsay cedex, France}
\email{eric.charron@u-psud.fr}

\date{\today}

\begin{abstract}
We extend the model of exciton-plasmon materials to include a ro-vibrational structure of molecules using wave-packet propagations on electronic potential energy surfaces. The new model replaces conventional two-level emitters with more complex molecules allowing to examine the influence of alignment and vibrational dynamics on strong coupling with surface plasmon-polaritons. We apply the model to a hybrid system comprising a thin layer of molecules placed on top of a periodic array of slits. Rigorous simulations are performed for two types of molecular systems described by vibrational bound-bound and bound-continuum electronic transitions. Calculations reveal new features in transmission, reflection and absorption  spectra including the observation of significantly higher values of the Rabi splitting and vibrational patterns clearly seen in the corresponding spectra. We also examine the influence of anisotropic initial conditions on optical properties of hybrid materials demonstrating that the optical response of the system is significantly affected by an initial pre-alignment of the molecules. Our work demonstrates that pre-aligned molecules could serve as an efficient probe for the sub-diffraction characterization of the near-field near metal interfaces.
\end{abstract}


\maketitle

%=====================
\section{Introduction}
%=====================

Metal nanostructures such as periodic arrays of holes\cite{RevModPhys.79.1267} or nanoparticles of various shapes\cite{doi:10.1021/cr200061k} offer a wide variety of optical properties to explore and utilize\cite{Stockman:11}. Owing to the coupling of incident resonant radiation to collective oscillations of conductive electrons at the interface such systems exhibit resonant absorption/scattering in the visible. The electromagnetic modes associated with density charge oscillations referred to as surface plasmon-polaritons (SPP) have unique properties including high spatial localization and coherence\cite{Gramotnev:2014aa}. The subdiffraction electromagnetic (EM) field localization of SPP modes can be utilized to control individual quantum emitters\cite{PhysRevApplied.1.014007} and their aggregates\cite{doi:10.1021/nn4054528}.

A new type of nanomaterials, namely hybrid materials recently became a hot topic of research\cite{0034-4885-78-1-013901} due to various possible applications ranging from fundamental understanding of light-matter interaction to EM energy control at the nanoscale\cite{Sukharev:2016aa}. When the coupling strength between quantum emitters surpasses any damping rates in a hybrid system at zero-detuning conditions the optical response exhibits a unique signature of the strong coupling, \emph{i.e.} Rabi splitting of a given resonant mode forming two new hybrid states, namely upper and lower polaritons. Conventional models describing the optical response of quantum emitters usually treat these emitters as simple interacting 2-level systems\cite{3167407,PhysRevA.84.043802} or as coupled Lorentz oscillators\cite{doi:10.1021/nl4014887}. Such models, for instance, have been used recently to predict and interpret a new phenomenon of transparency induced by a strong dipole-dipole interactions\cite{PhysRevLett.113.163603,PhysRevA.91.043835}. While such approaches include a wide range of experimentally observed phenomena they obviously neglect vibrational and rotational degrees of freedom exhibited by molecular emitters. In principle, contributions from vibrational degrees of freedom of molecular aggregates to the optical response of exciton-plasmon materials in the visible can be neglected due to different time scales of plasmons and phonons. This however does not include vertical electronic transitions (Frank-Condon transitions) between different electronic potential energy surfaces in each molecule, which may reveal a particular ro-vibrational structure of excited electronic states. The latter is obviously neglected in both approaches considering either 2-level emitters or Lorentz oscillators.

In this manuscript we extend the Maxwell-Bloch model\cite{PhysRevA.84.043802} and explicitly include ro-vibrational levels within each electronic state for each emitter. Using the example of interacting diatomic molecules with pre-defined potential energy curves we couple the ro-vibrational dynamics to Maxwell's equations in a self-consistent manner. In the next section we introduce our model. Next the model is applied to scrutinize linear optical properties of periodic arrays of slits strongly coupled to interacting diatomic molecules.

%===============================
\section{Theoretical model}
\label{model}
%===============================

The interaction of EM radiation with molecules is considered within a quasi-classical model, in which EM fields are treated classically and the optical response of the molecular subsystem is described using quantum mechanics. In the model EM fields, $\vec{E}$ and $\vec{B}$, satisfy Maxwell's equations
   \begin{subequations}
   \label{Maxwell}
    \begin{align}
      \frac{\partial \vec{B}}{\partial t}  &=  -\nabla \times \vec{E}, \\
     \varepsilon_0\frac{\partial \vec{E}}{\partial t}  &=  \frac{1}{\mu_0 }\nabla \times \vec{B} - \vec{J}, 
    \end{align}
  \end{subequations}
where $\varepsilon_0$ and $\mu_0$ are the permittivity and the permeability of free space, respectively. The current source in the Ampere law (\ref{Maxwell}b), $\vec{J}$, corresponds to either the current density in spatial regions occupied by the metal, Eq.(\ref{Drude_J}) below, or the macroscopic polarization current, $\vec{J}=\partial\vec{P}/\partial t$, in space filled with molecules. 

With the knowledge of the electric field $\vec{E}(\vec{r},t)$ the molecular dynamics is described by solving the non-Hermitian Schr\"odinger equation that we have developed recently\cite{jcp1.4774056,1.4947140}. Here we recall the particular points of this model that are essential for understanding our forthcoming discussion. For a detailed description of the model, see Refs. [\onlinecite{jcp1.4774056}] and [\onlinecite{1.4947140}]. This model was designed to treat the electronic excitation of neutral diatomic molecules, including the induced ro-vibrational dynamics. More specifically, we have chosen the particular case of a $^1\Sigma_g^+$ ground state and a $^1\Sigma_u^+$ excited electronic state. This corresponds, for instance, to the case of the ground and first excited states of alcali molecules such as Li$_2$ or Na$_2$. We follow the electronic dynamics and the nuclear motion by expanding the total molecular wave function as
\begin{equation}
\Psi(\mathbi{r},\mathbi{R},t) = \chi_g(\mathbi{R},t) \Phi_g(\mathbi{r},R) + \chi_e(\mathbi{R},t) \Phi_e(\mathbi{r},R)
\end{equation}
where $\Phi_g(\mathbi{r},R)$ and $\Phi_e(\mathbi{r},R)$ denote the electronic wave functions associated with the ground and excited electronic states. The electron coordinates are denoted by $\mathbi{r}$. $\mathbi{R}$ denotes the internuclear coordinate and we adopt Hund's case (b) representation for the description of the electronic wave functions in the molecular frame\cite{PhysRevA.74.033407}. The ro-vibrational time-dependent wave functions $\chi_g(\mathbi{R},t)$ and $\chi_e(\mathbi{R},t)$ are expanded on a limited set of normalized Wigner rotation matrices as\cite{PhysRevA.49.R641}
\begin{subequations}
	\begin{eqnarray}
	\chi_g(\mathbi{R},t) & = & \sum_{N,M} \chi^{(g)}_{N,M}(R,t) \, D^{N^{\,*}}_{M,0}(\hat{R}),\\
	\chi_e(\mathbi{R},t) & = & \sum_{N,M} \chi^{(e)}_{N,M}(R,t) \, D^{N^{\,*}}_{M,0}(\hat{R})\,,
	\end{eqnarray}
\end{subequations}
where $N$ denotes the molecular rotational quantum number while $M$ denotes its projection on a fixed axis of the laboratory frame. Introducing these expansions in the time-dependent Schr\"odinger equation describing the molecule-field interaction and projecting onto the electronic and rotational basis functions yields, in the dipole approximation, a set of coupled differential equations\cite{jcp1.4774056} for the nuclear wave packets $\chi^{(g)}_{N,M}(R,t)$ and $\chi^{(e)}_{N,M}(R,t)$ that we solve using the short-time split-operator method\cite{FEIT1982412}. To integrate the corresponding differential equations we need to define the potential energy curves $V_{g}(R)$ and $V_{e}(R)$ associated with the ground and excited electronic states of the molecule and the matrix elements which couple the nuclear wave packets evolving on these electronic potential curves. Such matrix elements are given in Ref. [\onlinecite{jcp1.4774056}] and for the potential curves we use Morse potentials with $V(R) = T + D ( 1 - \exp[ - a ( R - R_e ) ] )^2 - D$ for a bound state and $V(R) = T + 2 D \exp[ - a ( R - R_e )  ]$ for a dissociative state. The parameters used in our numerical simulations in the case of bound-bound transitions and in the case of bound-continuum transitions are given in Table\,\ref{Table_Pot}. These parameters have been chosen such that the absorption spectra of the molecules lie in the energy domain of the SPP resonance.

We assume that the molecules are prepared initially in the ground vibrational level of the ground electronic state. In terms of rotation, the molecules are assumed to be initially isotropically distributed (with $N=M=0$) or pre-aligned (see Section\,\ref{slits}). The essence of our non-Hermitian Schr\"odinger approach lies in the introduction of time-dependent gain and decay rates\cite{jcp1.4774056,1.4947140} for the ground and excited electronic states. These rates can be viewed as the positive and negative imaginary parts of the ground and excited states energies which are simple functions\cite{jcp1.4774056,1.4947140} that depend on the nonradiative decay rate of the excited state, the pure dephasing rate of the molecular system, and the population difference between the ground and excited electronic states. One can show that in weak fields with such time-dependent gain and decay rates the non-Hermitian Schr\"odinger approach\cite{jcp1.4774056} is strictly equivalent to the Liouville-von Neumann approach\cite{PhysRevA.84.043802} but it is numerically much more efficient because one propagates wave functions instead of density matrices.

\begin{table}[ht]
	\caption{Parameters used in the definition of the potential curves.}
	\begin{ruledtabular}
		(a) Bound-Bound Transitions (BB Model)\\[0.2cm]
		\begin{tabular}{l c c c c}
			                      & $T$ (eV) & $D$ (eV) & $a$ (a.u.) & $R_e$ (a.u.) \\ [0.5ex]
			\hline \noalign{\vskip 1.0ex}
			Ground State  (bound) & 0        & 1.050    & 0.457917   & 5.06         \\ [0.5ex]
			Excited State (bound) & 1.867    & 1.160    & 0.317763   & 5.89         \\ [0.5ex]
		\end{tabular}
	\end{ruledtabular}
	\begin{ruledtabular}
		\\
		(b) Bound-Continuum Transitions (BC Model)\\[0.2cm]
		\begin{tabular}{l c c c c}
			                             & $T$ (eV) & $D$ (eV) & $a$ (a.u.) & $R_e$ (a.u.) \\ [0.5ex]
			\hline \noalign{\vskip 1.0ex}
			Ground State  (bound)        & 0        & 1.050    & 0.457917   & 5.06         \\ [0.5ex]
			Excited State (dissociative) & 0        & 0.575    & 0.332252   & 3.98         \\ [0.5ex]
		\end{tabular}
	\end{ruledtabular}
	\label{Table_Pot}
\end{table}

%====================================================
\section{Application to periodic arrays of slits}
\label{slits}
%====================================================

\begin{figure}[t!]
\begin{center}
\includegraphics[width=\textwidth]{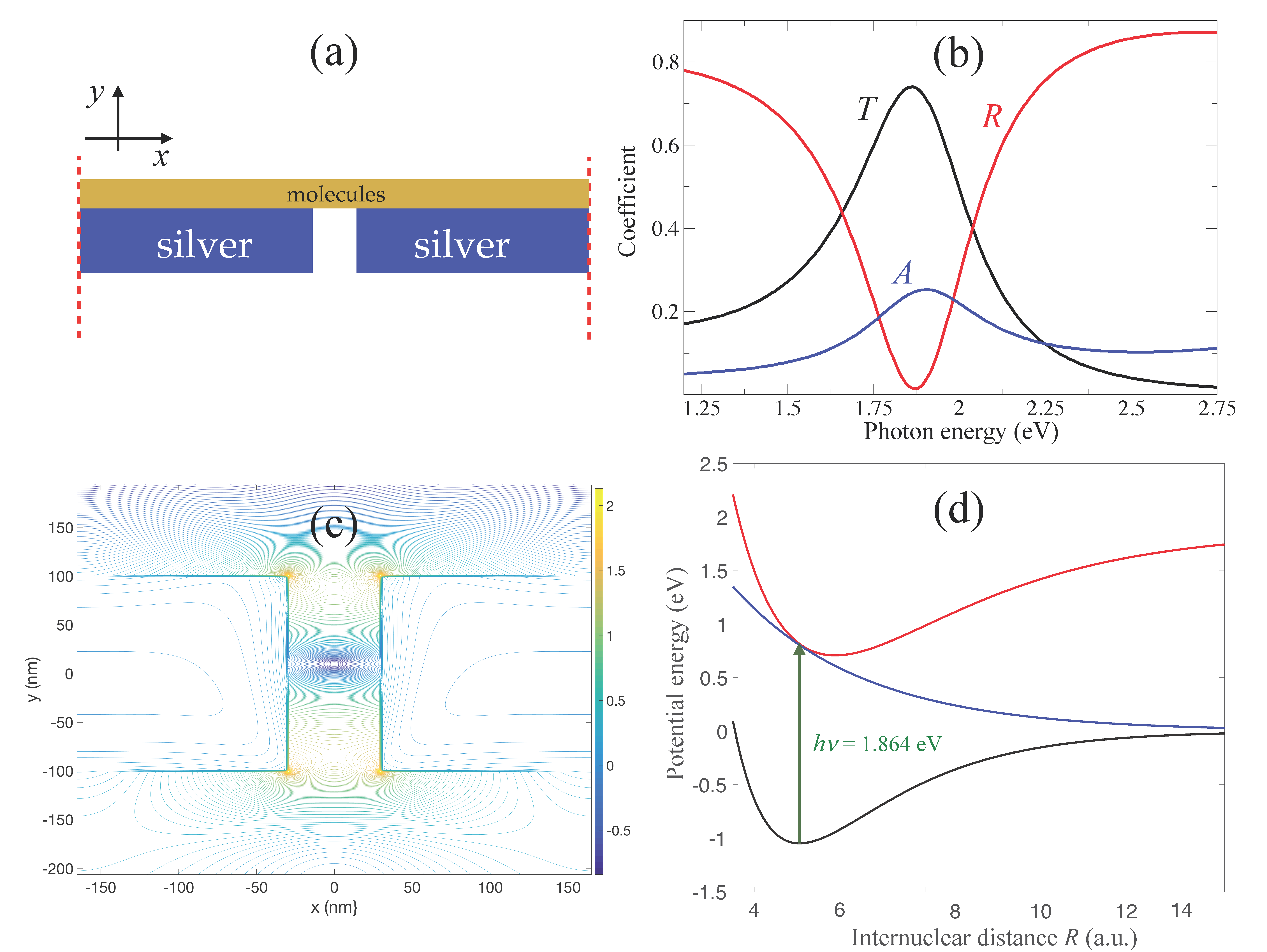}
\caption{(Color online) Panel (a) shows a schematic setup of the simulations with vertical red dashed lines indicating period boundaries. The thickness of the silver film is fixed at $200$\,nm and the slits' width is $60$\,nm. The external probe is launched from the top propagating in the negative $y$-direction and it is horizontally polarized. Panel (b) shows transmission ($T$, black line), reflection ($R$, red line), and absorption ($A$, blue line) as functions of the incident photon energy for the bare slit array at the period of $330$\,nm. Panel (c) shows the EM intensity distribution corresponding to the photon energy $1.864$\,eV (maximum of the transmission in panel (b)). The intensity is plotted in logarithmic scale and is normalized to the intensity of the incident field. Panel (d) shows the potential energy curves used for the model molecular system as functions of the internuclear distance $R$. The black line shows the ground electronic state, the red line shows the bound excited electronic state, and the blue line shows the dissociative excited state.}
\label{fig1}
\end{center}
\end{figure}

To understand how the ro-vibrational dynamics of the interacting molecules affect the optical properties of exciton-plasmon materials we consider a periodic array of slits with a thin molecular layer placed on top of the metal as shown in Fig.\,\ref{fig1}a. The dispersion of the metal is taken into account via the conventional Drude model with the dielectric constant, $\varepsilon\left(\omega\right)$, in the form
 \begin{equation}
\label{Drude_epsilon}
 \varepsilon\left(\omega\right)=\varepsilon_r-\frac{\omega_p^2}{\omega^2-i\Gamma\omega},
\end{equation}
where $\Gamma$ is the damping parameter, $\omega_p$ is the bulk plasma frequency, and $\varepsilon_r$ is the high-frequency limit of the dielectric constant. For the range of frequencies considered in this work the following set of parameters was chosen to represent silver: $\varepsilon_r=8.926$, $\omega_p=11.585$\,eV, and $\Gamma=0.203$\,eV\cite{PhysRevB.68.045415}. The corresponding current density in the metal region is evaluated according to the time-dependent equation\cite{taflove2005computational}
 \begin{equation}
\label{Drude_J}
 \frac{\partial\vec{J}}{\partial t}+\Gamma\vec{J}=\varepsilon_0\omega_p^2\vec{E}.
\end{equation}

The system is excited by a plane wave at normal incidence propagating in the negative $y$-direction. We employ the short pulse method\cite{PhysRevA.84.043802} and evaluate refection and transmission in the far-field zone as functions of the  incident photon energy. The thickness of the metal is fixed at $200$\,nm and the slits' width is set to $60$\,nm. The slits' period is varied between $200$\,nm and $500$\,nm. It is seen in Fig.\,\ref{fig1}b that a bare slit array exhibits a sharp resonance near $1.864$\,eV for the period of $330$\,nm. The maximum of transmission at this energy indicates that this peak is a SPP resonance\cite{0034-4885-72-6-064401}. The steady-state EM intensity distribution at this frequency is shown in Fig.\,\ref{fig1}c in logarithmic scale. The field is highly localized near the sharp edges of the slits extending both into the slits and into outer surfaces on the input and output sides. It should be noted that the transmission reaches almost $80\%$ at the SPP resonance with a relatively high $Q$-factor of $4$ for an SPP mode. The high $Q$-factor and the fact that the local field enhancement is quite significant promises a noticeable EM coupling to molecules placed in close vicinity of the slits.

Our choice of particular parameters describing the diatomic molecules is somewhat arbitrary as our major goal in this paper is to investigate the coupling between molecules and SPP waves on a qualitative level. Both the high sensitivity of SPP modes to various geometrical parameters\cite{Stockman:11} and outstanding progress in nanofabrication techniques\cite{maradudin2014modern} would allow one to construct a plasmonic system capable of coupling to a molecule of one's choice. We consider two frequently used models for a molecule depicted in Fig.\,\ref{fig1}d with a bound ground state and either bound or dissociative excited electronic state. We refer to the first model of bound-bound transitions as the BB-model and to the second model with bound-continuum transitions as the BC-model. In our simulations the ground state is described by a Morse potential with a minimum located near the internuclear distance $R=5.06$\,a.u. The bound excited state has a minimum at $R=5.89$\,a.u. Both bound and dissociative excited states are chosen such that the vertical transition from the ground state has an energy of $1.864$\,eV, corresponding to the SPP resonance of a slit array at the period of $330$\,nm.

\begin{figure}[t!]
\begin{center}
\includegraphics[width=\textwidth]{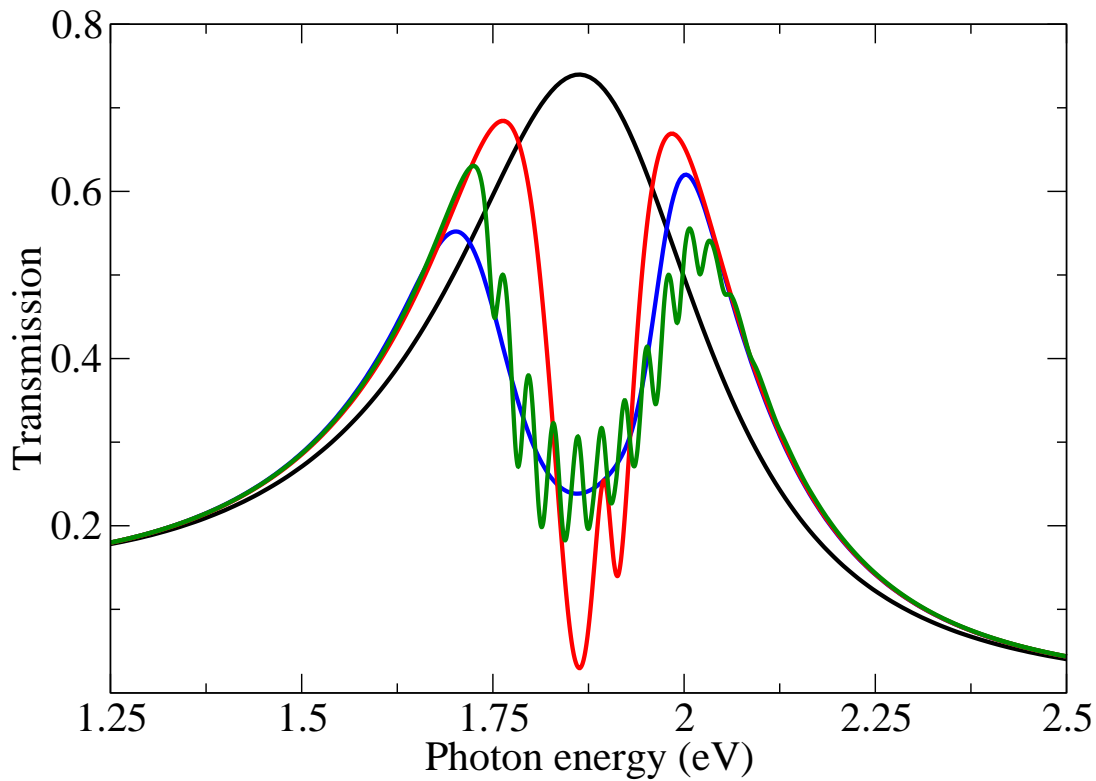}
\caption{(Color online) Transmission as a function of the incident photon energy. The black line corresponds to the bare slit array. The red line shows data for the $2$-level model. The green line shows the BB-model and the blue line shows the BC-model. The period of the array is $330$\,nm. The transition dipole for the $2$-level model is $7.63$\,Debye and the molecular transition energy is $1.864$\,eV. Other parameters for all molecular models are: the molecular concentration is $5\times10^{26}$\,m$^{-3}$, the nonradiative decay time is $1$\,ps, the pure dephasing time is $600$\,fs.}
\label{fig2}
\end{center}
\end{figure}

Fig.\,\ref{fig2} shows a direct comparison between the transmission obtained with bare slits (black line) and the transmission obtained with three molecular models: (a) the conventional model with interacting $2$-level emitters with no ro-vibrational dynamics included (red line); (b) the BB-model (green line); (c) the BC-model (blue line). Firstly, it can easily be seen that the transmission spectra significantly differ for all three models, the $2$-level model resulting in the smallest Rabi splitting ($220$\,meV). It should be noted that the small resonance observed at $1.895$\,eV for the $2$-level model is the collective mode, physics of which is discussed elsewhere\cite{PhysRevLett.109.073002,fauche2016plasmonic}. We note that the collective mode also appears in the transmission spectra obtained with the BB- and BC-models but at higher molecular concentrations (not shown here). Secondly, the BB-model clearly shows the signature of vibrational states in the transmission spectrum. Further analysis confirms that the sharp peaks seen in the transmission spectrum for the BB-model directly correspond to the vibrational states in the excited electronic state. Their relative weight in the spectrum is a signature of their Frank-Condon factors and of the strength of the local EM field at a given incident frequency. The visibility of the vibrational states in the spectra is highly dependent upon the pure dephasing rate phenomenologically introduced in our model (see Section\,\ref{model}). For the parameters considered here, the vibrational states become indistinguishable (their widths exceed the energy spacing) for dephasing rates above $21$\,meV, \emph{i.e.} for a characteristic time less than about $200$\,fs, corresponding approximately to the lifetime of the SPP mode. We thus conclude that the vibrational ladder seen in the transmission may in principle be observed at room temperatures for plasmonics systems with modes with high Q-factors,\cite{Lilley:15} \emph{i.e.} long lifetimes. The largest Rabi splitting among the three models is exhibited by the BC-model resulting in about $300$\,meV. The latter is readily explained by comparing the Frank-Condon factors for the BB and BC models (not shown here) as the dissociative state accommodates a slightly wider range of energies accessible to a one photon transition from the ground state. Upon examining reflection/absorption spectra (not shown) we also observe the vibrational patterns with the same characteristics as in transmission.

\begin{figure}[t!]
\begin{center}
\includegraphics[width=\textwidth]{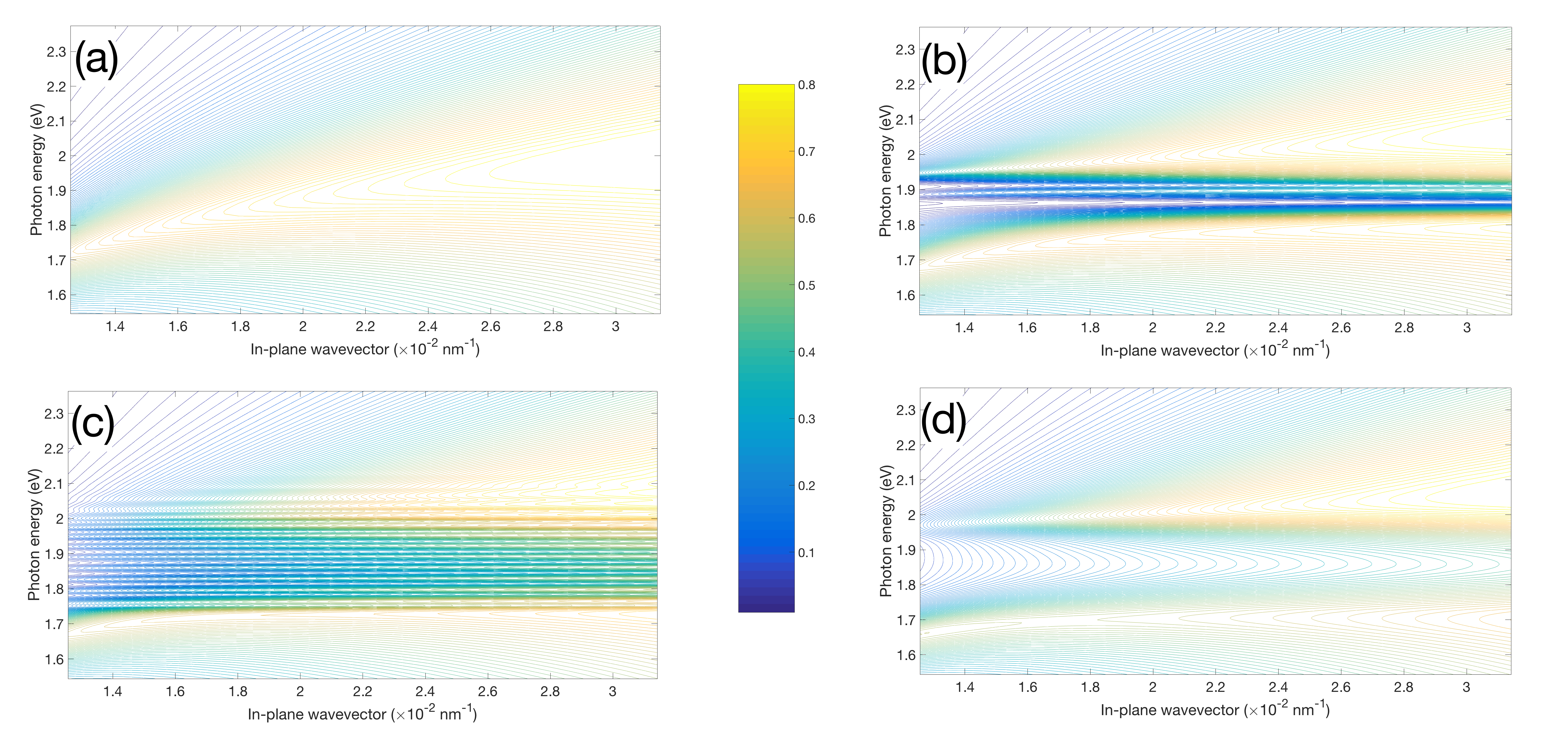}
\caption{(Color online) Transmission spectra as functions of the incident photon energy and in-plane $k$-vector (defined as $2\pi/$period) for (a) the bare slit array, (b) slit array covered by molecules treated as interacting $2$-level systems with no ro-vibrational degrees of freedom, (c) slit array covered by molecules described by the BB model, and (d) slit array covered by molecules described by the BC model. The molecular transition energy for the $2$-level model is set at $1.864$\,eV. The molecular density for all models is $5\times 10^{26}$\,m$^{-3}$, the non-radiative decay time is $1$\,ps, and the pure dephasing rate is $600$\,fs.}
\label{fig3}
\end{center}
\end{figure}

To examine the behavior of the optical coupling between the molecules and the SPP mode we perform a series of simulations gradually varying the period of the array under normal incidence and calculating the transmission spectra for all three molecular models. The results of these simulations are presented in Fig.\,\ref{fig3}. The dispersion of the SPP mode seen in Fig.\,\ref{fig3}a shows that the mode becomes broader at shorter periods (or higher $k$-vectors) and that its resonant energy shifts to the blue. We can explain such a behavior if we recall that shorter periods essentially mean narrower metal stripes with shorter transverse resonant wavelengths leading to higher losses and higher resonant energies. The transmission obtained with $2$-level molecules coupled to the SPP is shown in Fig.\,\ref{fig3}b. It demonstrates two expected features: (a) an avoided crossing is observed when the SPP mode passes through the molecular transition energy; (b) the collective resonance seen in Fig.\,\ref{fig2} as a sharp peak at $1.895$\,eV is nearly dispersionless in agreement with earlier findings.\cite{PhysRevLett.109.073002,fauche2016plasmonic} A noticeably reacher physics is observed in Figs.\;\ref{fig3}c and \ref{fig3}d as we examine how the BB and BC models influence the coupling with the SPP mode. The signature of the vibrational spectrum previously discussed for the BB-model is clearly visible in Fig.\,\ref{fig3}c as a distinct set of resonances in the spectral region between $1.75$\,eV and $2.10$\,eV. Moreover one should note how high energy vibrational states appear in the transmission at higher wave vectors as the SPP mode shifts to the blue. Results for the BC-model are shown in Fig.\,\ref{fig3}d. Similarly to the previous model we observe the avoided crossing behavior in the transmission for this dissociative model as well. The Rabi splitting is readily higher as it was pointed out above, leading to a higher band gap observed in the transmission.

Up to this point our simulations were relying on isotropic initial conditions, assuming that the molecules were initially in the ground rotational level $N=M=0$. It is however possible to pre-align molecules along $x$ or $y$ and to explore how the optical properties of the hybrid system vary with the direction and with the degree of alignment. For this study, we will consider molecules whose initial angular distribution verifies $P(\theta) \propto \cos^{2n}(\theta)$ where $\theta$ denotes either the angle between the molecular axis and the $x$-axis or the angle between the molecular axis and the $y$-axis. Using this simple convention, it is possible to explore both the influence of the alignment direction ($x$ or $y$) and the influence of the degree of alignment, controlled by the alignment parameter $n$, higher values of $n$ yielding stronger alignment.

\begin{figure}[t!]
	\begin{center}
		\includegraphics[width=0.95\textwidth]{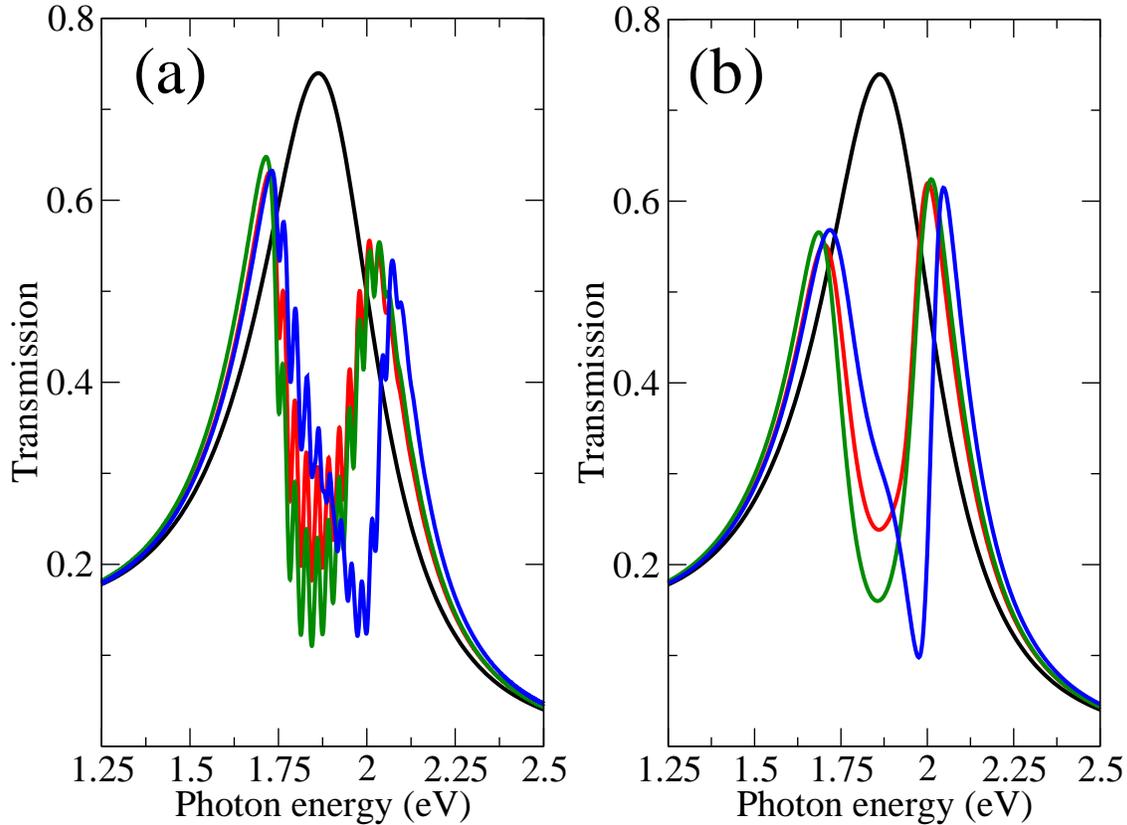}
		\caption{(Color online) Transmission as a function of the incident photon energy. Panel (a) shows simulations for the BB model. Panel (b) corresponds to simulations performed using the BC model. In both panels transmission for the bare array of slits is indicated as a black line. Red lines correspond to isotropic initial conditions, green lines are data for molecules pre-aligned along $x$ with the degree of alignment $n=3$. Blue lines are results of simulations for molecules pre-aligned along $y$, with the same degree of alignment $n=3$. The period of the array in all simulations is $330$\,nm. Other parameters for all molecular models are: a molecular concentration of $5\times10^{26}$\,m$^{-3}$, a non-radiative decay time of $1$\,ps and a pure dephasing time is $600$\,fs.}
		\label{fig4}
	\end{center}
\end{figure}

Fig.\,\ref{fig4} explores two anisotropic initial conditions and compares the BB and BC models. First we note that when the molecules are initially aligned in the direction transverse to the propagation of the incident radiation ($x$-axis in Fig.\,\ref{fig1}a) the transmission spectra for both models (green lines in Fig.\,\ref{fig4}) are similar to those we obtain with isotropic initial conditions (red lines in Fig.\,\ref{fig4}). The values of the Rabi splitting for molecules pre-aligned along $x$ are similar to the ones obtained in the isotropic case. However in both the bound-bound and bound-continuum cases we observe lower transmission at the SPP energy of $1.864$\,eV corresponding to the minimum in the transmission spectra, an indication of a stronger coupling between the molecules and the SPP resonance for molecules pre-aligned along the $x$-axis. Another anisotropic case considered is with molecules initially aligned in the direction of the incident field propagation ($y$-axis in Fig.\,\ref{fig1}a). A clear distinction between the two anisotropic initial conditions is observed. The $y$-orientation pushes the minimum in transmission to higher energies and lower values. Moreover the energies of the upper and lower polaritonic states also experience a blue shift. It is important to note that the calculated transmission spectra are barely dependent on the degree of alignment $n$, \emph{i.e.} a similar trend is seen for $n=1$ and for higher values of $n$. In order to elucidate the observed differences between $x$ and $y$ pre-aligned molecules we examined the local EM field of the resonant SPP mode. Simulations reveal the fact that the local field influencing the molecular layer on the input side is predominantly $y$-polarized while the field inside the slits is mainly polarized along $x$. Additional simulations with molecules placed inside slits instead of a flat film on top of the slit array further confirm that the polarization of the SPP field is a major factor responsible for the observed differences in Fig.\,\ref{fig4}. As a consequence, it appears that pre-aligned linear molecules could serve in the future as an efficient probe for the characterization of the SPP field polarization.

%================================
\section{Summary and conclusions}
\label{sec:summary}
%================================

In summary we extended the conventional model of exciton-plasmon materials based on simple interacting two-level emitters by explicitly including the ro-vibrational structure of diatomic molecules using time-dependent wave packet propagation on electronic potential energy surfaces. The new model allows to examine the influence of alignment and vibrational dynamics on strong coupling with surface plasmon-polaritons. We apply our model to periodic slit arrays coupled to a thin layer of interacting molecules. Rigorous simulations were performed for two types of molecular systems described by bound-bound and bound-continuum (dissociative) transitions. Our calculations show new distinct features in the transmission, reflection, and absorption spectra, including the observation of significantly higher values of the Rabi splitting and the vibrational ladder clearly seen in the corresponding spectra. We also examined the influence of anisotropic initial conditions on the optical properties of hybrid materials demonstrating that initially pre-aligned molecules significantly affect the optical response of the system. Our future work is to examine pump-probe experiments and to explore how initially excited ro-vibrational wave packets propagating on excited electronic surfaces influence the optics of various hybrid systems including periodic arrays of holes and core-shell nanoparticles.

%==========================
\section*{Acknowledgements}
%==========================

M.S. is grateful to the Universit\'e Paris-Sud (Orsay) for the financial support through an invited Professor position in 2016. M.S. would also like to acknowledge financial support by the Air Force Office of Scientific Research under grant No. FA9550-15-1-0189 and Binational Science Foundation under grant No. 2014113. E.C. is grateful for the use of the computing center GMPCS of the LUMAT federation (FR LUMAT 2764). E.C. would also like to acknowledge financial support of the EU (Project ITN-264951) and of the Partenariat Hubert Curien (PHC) Procope mobility program 2016.

\bibliographystyle{apsrev4-1}

\end{document}